\documentclass[12pt]{iopart}
\usepackage{graphicx}

\expandafter\let\csname equation*\endcsname\relax
\expandafter\let\csname endequation*\endcsname\relax

\usepackage{amsthm}
\usepackage{amsmath}
\usepackage{iopams}
\usepackage{enumitem}
\usepackage{placeins} 

\newcommand{\CA}{{\cal A}}
\newcommand{\CL}{{\cal L}}
\newcommand{\CC}{{\cal C}}

\newcommand{\bfq}{\mathbf{q}}

\newcommand{\CNtxt}{\hmm{c}losest neighbour}
\newcommand{\CN}{\text{CN}}
\newcommand{\SCN}{\text{2CN}}

\newcommand{\kCNtxt}{$k$-th closest neighbour}
\newcommand{\kCN}{\text{$k$CN}}

\newcommand{\FNtxt}{\hmm{f}arther neighbour}
\newcommand{\FN}{\text{FN}}

\newcommand\hmm[1]{\ifnum\ifhmode\spacefactor\else2000\fi>1000 \uppercase{#1}\else#1\fi}

\newcommand{\SCNtxt}{second closest neighbour}

\newcommand{\ud}{\text{d}}
\newcommand{\ui}{\text{i}}
\newcommand{\ue}{\text{e}}

\newcommand{\ld}{\lambda}

\usepackage[compress, nosort]{cite}  

\begin{document}
\title{Ordered level spacing probability densities}

\author{Shashi C. L. Srivastava$^{1,2,3}$,
        Arul Lakshminarayan$^4$,\\
        Steven Tomsovic$^5$,
        Arnd B\"acker$^{3,6}$}

\address{$^{1}$Variable Energy Cyclotron Centre, Kolkata 700064, India.}
\address{$^2$Homi Bhabha National Institute, Training School Complex,
Anushaktinagar, Mumbai - 400085, India}
\address{$^3$Max-Planck-Institut f\"{u}r Physik komplexer Systeme,
N\"{o}thnitzer Strasse 38,\\
 01187 Dresden, Germany}
\address{$^{4}$Department of Physics, Indian Institute of Technology
Madras, Chennai, India~600036}

\address{$^{5}$Department of Physics and Astronomy, Washington State
University, Pullman, WA~99164-2814}

\address{$^6$Technische Universit\"{a}t Dresden,\\
          Institut f\"{u}r Theoretische Physik and Center for Dynamics,\\
          01062 Dresden, Germany}


\begin{abstract}
Spectral statistics of quantum systems have been
studied in detail using the nearest neighbour level spacings,
which for generic chaotic systems follows random matrix theory predictions.
In this work, the probability density of the \CNtxt{} and \FNtxt{}
spacings from a given level are introduced.  Analytical predictions are derived using a $3\times 3$ matrix model.  The \CNtxt{} density is generalized to the \kCNtxt{}
spacing density, which allows for investigating long-range correlations.
For larger $k$ the probability density of \kCNtxt{} spacings is well described by a Gaussian.
Using  these \kCNtxt{} spacings we propose the ratio of the \CNtxt{} to
the \SCNtxt{} as an alternative to the ratio of successive spacings.
For a Poissonian spectrum the density of the ratio is flat, whereas for
the three Gaussian ensembles repulsion at small values is found.
The ordered spacing statistics and their ratio
are numerically studied for the integrable circle billiard,
the chaotic cardioid billiard, the standard map and
the zeroes of the Riemann zeta function.  Very good agreement
with the predictions is found.

\end{abstract}

\vspace{2pc}

\maketitle

\section{Introduction}

Random matrix theory (RMT) describes quite successfully the
statistical properties of complex systems spectra of various origins,
such as nuclear and condensed matter systems, and has applications in
mathematics as well \cite{GuhMueWei1998,RMTScholarpedia}.
\emph{Consecutive level spacing statistics}\footnote{
Commonly this is referred to as
\emph{nearest neighbour spacing distribution}.
However, for the purposes of this paper this would lead to confusion;
to avoid this will use \emph{consecutive level spacing probability densities}
throughout the text.}
belong to one of the simplest classes of measures
and has been used to differentiate between
the spectra of integrable and chaotic origin.
Generic integrable systems are expected to follow
Poissonian statistics \cite{BerTab1977}, whereas
the stastistics of classically strongly
chaotic systems are described by the corresponding
RMT densities \cite{BohGiaSch1984}.
Other important commonly used fluctuation measures are given by the number variance
and spectral rigidity \cite{DysMeh1963,BroFloFreMelPanWon1981,Ber1985,Ber1988}.
Their universal behaviours create the opportunity to search for deviations that reveal interesting
physics arising in physical systems, e.g.~localization or other effects.

In this paper a general class of spacing statistics is introduced,
which is based on the ordering of the levels according
to the increasing distance from a given level as opposed to consecutive levels.
The simplest are the densities of the spacings of
the \emph{\CNtxt{}} and that of the \emph{\FNtxt{}}.
One motivation for this comes
from applications in the context of perturbation theory
\cite{LakSriKetBaeTom2016,TomLakSriBae2018},
where it is of importance to know the statistics
of the distance of a given level to the closer neighbour (\CN) of its two nearest neigbors.
The reason is that this gives the strongest contribution
due to a smaller energy denominator.
The \CNtxt{} spacing density has already
been discussed for the Gaussian unitary ensemble
in Refs.~\cite{ForOdl1996, HerOngUsaMatBar2007}.
The complement of the \CNtxt{} is the
\FNtxt{} (\FN).  In general, there is the set of {\kCNtxt} (\kCN) and
these determine the most significant contributions in the context of perturbation theory and allow for testing longer range correlations in the spectra.

For the \CNtxt{}  and  \FNtxt{} spacing densities
random matrix predictions are derived for the
Gaussian orthogonal (GOE),
Gaussian unitary (GUE), and
Gaussian symplectic (GSE) ensembles.
The results for Poissonian spectra describing the case of systems with integrable
dynamics are also given.
These predictions are successfully compared
with results for the densities
for several disparate systems: the integrable circle billiard, completely chaotic case of the cardioid billiard, the kicked (Floquet) system of the  standard map and the statistics of the zeros of the Riemann zeta function, which is known to have properties of the GUE spectra.

One drawback of level spacing densities, especially for many-body systems,
is the requirement of unfolding spectra to unit mean spacing.
Unfolding works best with an analytical expression for the smoothed density of eigenvalues,
such as the Weyl formula for billiards \cite{BalHil1976}, which is not always available. To overcome this difficulty, the ratio of the closer neighbour spacing to the farther neighbour spacing ($r=$ \CN/\FN) was introduced in Ref.~\cite{OgaHus2007}. While this ratio has quickly become popular in the context of many body localization as indicator of chaotic behaviour
\cite{AtaBogGirRou2013, AtaBogGirVivViv2013, ChaDeoKot2014, AleRig2014,
KheChaKimSon2014} and has been studied for various RMT ensembles, the spacings {\CN} and {\FN} themselves are neglected.
Based on the \CNtxt{} and \SCNtxt{} spacings we propose an alternative, i.e.~the statistics of their ratio  $r^\CN{}$, defined in Eq.~\eqref{eq:r-i-CN},
which also does not require an unfolding. As the second closest neighbour
could be the farther neighbour or the next-neighbour level
in the direction of the closest, this ratio will be in general larger than $r$.

\section{Ordered spacing probability densities}

\begin{figure}[b]
  \begin{center}
    \includegraphics{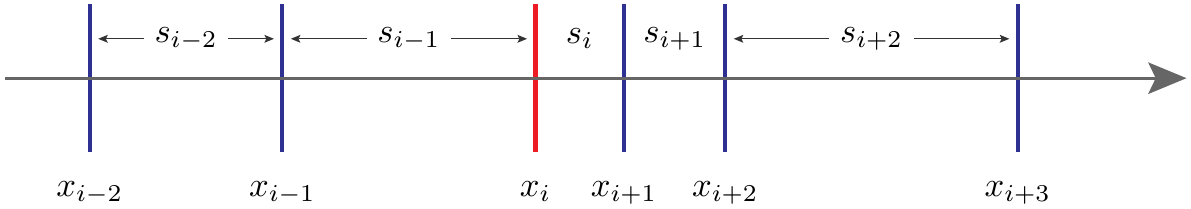}
  \end{center}
  \caption{\label{fig:sketch-x-i}
           Sketch of the neighbours of $x_i$.           %
           In this example $x_{i+1}$ is the \emph{\CNtxt} (\CN),
           while $x_{i-1}$ is the \emph{\FNtxt} (\FN).
           The \emph{\SCNtxt} (\SCN) is $x_{i+2}$,
           as $x_i$ is closer to $x_{i+2}$ than to $x_{i-1}$.
           Consecutive level spacing probability density is based on
           the $s_i = x_{i+1} - x_i$.
           }
\end{figure}

Consider a sequence of levels $\{x_i\}$ which are assumed
to be ordered, i.e.\ $x_{i+1} \ge x_i$,
and unfolded \cite{Por1965,BohGiaSch1984,BohGiaSch1984b};
see Fig.~\ref{fig:sketch-x-i}.
The consecutive level spacings $s_i = x_{i+1} - x_i$ have unit
average. The following types of ordered spacings will be studied:
\begin{itemize}
  \item \emph{\CNtxt} (\CN) spacing
        $
           s^{\CN}_i = \min(x_{i+1} - x_i, x_i - x_{i-1}),
        $
  \item \emph{\FNtxt} (\FN) spacing
         $
           s^{\FN}_i = \max(x_{i+1} - x_i, x_i - x_{i-1}),
         $
  \item \emph{\kCNtxt} (\kCN) spacing,
        $s^{\kCN}_i$, is obtained from the set of distances
\begin{equation*}
  \{ x_i - x_{i-k}, ..., x_i - x_{i-1};\
     x_{i+1} - x_i,  ..., x_{i+k} - x_i \},  \qquad k > 0,
\end{equation*}
by determining the $k$-th element after sorting the set.
For $k=1$ this reduces to the \emph{\CNtxt} (\CN)
and for $k=2$ to  the \emph{\SCNtxt} (\SCN) spacings.  The ordering operation is equivalent to cutting a spectrum at the $i^{th}$ level and reflecting its left or right portion to the opposite side.
\end{itemize}
Random matrix approximations for the probability density
of the {\FNtxt} and the {\CNtxt} are derived in sections~\ref{sec:fn} and \ref{sec:cn}
for the GOE, GUE, and GSE, (with Dyson indices $\beta=1,2,4$ respectively) based on a $3\times 3$ model, and for the Poissonian spectrum.
The \kCNtxt{} densities are discussed in section~\ref{sec:kcn}.
The results for the probability density of the \CNtxt{} are
in agreement with earlier results obtained for the GOE in
\cite{HerOngUsaMatBar2007}.  In section~\ref{sec:applications}
these results are compared with numerical computations for various dynamical systems and the zeros of the Riemann zeta function.

\subsection{Farther neighbour (\FN) spacings}
\label{sec:fn}

Start with the joint probability density (JPD) of the consecutive level spacings for the matrix ensemble under consideration. Despite being a highly
correlated sequence, the JPD can be written analytically by
performing a change of variable in the JPD of the eigenvalues. In principle,
all the spacings can be integrated excluding the successive two of interest.

An alternative approximate route, in the spirit of Wigner \cite{Wig1957},
is to consider minimal $3 \times 3$ random matrix ensembles.
Denote the JPD of the three ordered eigenvalues $\ld_1 \le \ld_2 \le \ld_3$ by
$P_{\beta}(\ld_1, \ld_2, \ld_3)$.  The \FN{} spacing density is the probability density of
max$(s_1, s_2)$ where $s_1= x_2-x_{1} = \mu_{\beta}(\ld_2-\ld_1)$
and $s_2=x_{3} - x_2=\mu_{\beta}(\lambda_3-\lambda_2)$
are the two successive spacings scaled by a factor $\mu_{\beta}$ in order to unfold to unit spacing.  Such a multiplicative scaling is sufficient as we are dealing only with three levels.
The following sequence gives the steps for calculating the \FN{} spacing density for the three Gaussian ensembles:
\begin{enumerate}
\item  The JPD of three ordered levels $\{\ld_1, \ld_2, \ld_3\}$ is given by \cite{Meh2004}
  \[
  P_{\beta}(\ld_1, \ld_2, \ld_3) =  \mathcal{N}_{\beta} \, | \ld_1-\ld_2|^\beta
|\ld_{1}-\ld_{3}|^\beta
  |\ld_2-\ld_{3}|^\beta \ue^{-\frac{\beta}{2} (\ld_{1}^2 + \ld_2^2 + \ld_{3}^2)} \Theta( \ld_2 - \ld_{1})
\Theta(\ld_{3} - \ld_2),
\]
with normalisation $\mathcal{N}_{\beta}$, and $\Theta(x)$ is the Heaviside step function.
\item The JPD of the spacings $s_1$, $s_2$ can be calculated
  from the JPD of the ordered eigenvalues as
  \begin{equation}
    \begin{aligned}
      P_{\beta}(s_1, s_2) &= \int \ud \ld_{1} \ud \ld_2 \ud \ld_{3} \,
  \delta(s_1-\mu_{\beta}(\ld_2-\ld_{1})) \delta(s_2-\mu_{\beta}(\ld_{3}-\ld_2))
     P_{\beta}(\ld_1, \ld_2, \ld_3) \\
   & = \tilde{{\cal{N}}_{\beta}} \, s_1^\beta s_2^\beta (s_1 + s_2)^\beta
 \exp\left(-\frac{\beta}{3 \mu_{\beta}^2}(s_1^2 + s_2^2 + s_1s_2)\right), \\
 \end{aligned}
\end{equation}
where $\tilde{{\cal{N}}}_{\beta}$ is a normalisation constant.
\item The constants $\tilde{{\cal{N}}_{\beta}}$ and $\mu_{\beta}$ are determined by normalization of $P_{\beta}(s_1,s_2)$ and requiring the unit average of its marginals, that is $\int_0^\infty s \, p_{\beta}(s) \ud s =1$, where $p_{\beta}(s) = \int_0^\infty P_{\beta}(s,
  s_2) \ud s_2$ is the consecutive level spacing density.
  This fact and the normalisation of $P_\beta(s_1, s_2)$ gives
for the JPD of the spacings
\begin{subequations} \label{eq:jpdfs1s2}
\begin{align}
	P^{\text{GOE}} (s_1, s_2) &=  4\sqrt{\frac {2} {3\pi}} \, a^{5/2}\, s_1
      s_2 (s_1 + s_2)
      \ue^{-\frac{2}{3}a(s_1^2 + s_1 s_2 + s_2^2)},
              \quad a
      =\frac{27}{8\pi},  \label{eq:jpdfs1s2_goe}\\
	P^{\text{GUE}} (s_1, s_2) &=  \frac{4}{\sqrt{3} \pi } \,b^4 \,s_1^2 s_2^2 (s_1+s_2)^2  \ue^{-\frac{2}{3}b (s_1^2+s_1s_2+s_2^2)},  \quad  b= \frac{729}{128\pi},\label{eq:jpdfs1s2_gue}\\
      P^{\text{GSE}} (s_1, s_2) &=  \frac{4}{45 \sqrt{3} \pi }\, c^7\, s_1^4 s_2^4 (s_1+s_2)^4
      \ue^{-\frac{2}{3}c (s_1^2+s_1s_2+s_2^2)},  \quad  c=\frac{3^{12}}{25\pi 2^{11}}.\label{eq:jpdfs1s2_gse}
\end{align}
\end{subequations}
Here, $P^{\text{GOE}}$ is $P_1$ etc., to be used where is the explicit abbreviation
of the ensemble is favored.  Also the scaling factors $\mu_i$ are related to the constants
$a,b,c$ as $a=1/2 \mu_1^2$, $b=1/\mu_2^2,$ and $c=2/\mu_4^2$.
Integrating one spacing results in the consecutive level spacing density $p_{\beta}(s)$.
For example, for the GOE this yields
\begin{equation}
  p_1(s) = \frac{a}{2 \pi } s \ue^{-\frac{2}{3}  a s^2} \left[\sqrt{6 \pi a }
    s-\pi  \ue^{\frac{a s^2}{6}} \left(a s^2-3\right) \text{erfc}\left(\sqrt{\frac{a} {6}}s\right)\right].  \label{eq:cls_goe}
\end{equation}
Although different from the Wigner surmise derived from two levels (see Eq.~(\ref{P(S)-GOE}) below), the  difference is negligible for nearly all applications.

\item The cumulative distribution of max $(s_1, s_2)$
  is defined as
  \begin{equation*}
   Q[\text{max}(s_1, s_2)< s] = Q[s_1 < s, s_2< s] = \int\limits_0^{s}
   \int\limits_0^{s} P(s_1, s_2) \ud s_1\ud s_2
 \end{equation*}
 from which the probability density is obtained
 via
 \begin{equation}
    P_{\text{\FN }}(s) = \frac{\ud}{\ud s} Q(s)= 2\int_0^s P(s_1,s) \ud s_1,
    \label{eq:fndist}
 \end{equation}
 where in the final step the symmetry in $s_i$ is used.

Therefore, the probability densities of the \FNtxt{} spacings are
  \begin{subequations} \label{eq:fnns}
   \begin{align}
	P^{\text{GOE}}_{\text{\FN}} (s) &=  \frac{a}{\pi } s \ue^{-2 a s^2}  \left[\pi  \ue^{\frac{3 a s^2}{2}}
        \left( a s^2-3 \right) \left\{ \text{erf} \left( \frac{\sqrt{a}
              s}{\sqrt{6}} \right) - \text{erf} \left( \sqrt{\frac{3}{2}}
            \sqrt{a} s \right) \right\} \right. \nonumber \\
            &\left.  \quad \quad
        \quad \quad \quad \quad +\sqrt{6 \pi }
        \sqrt{a} s \left(\ue^{\frac{4 a s^2}{3}}-3\right)\right],
        \label{eq:fnns_goe}\\
	P^{\text{GUE}}_{\text{\FN}} (s)  &= \frac{b^{3/2}}{4 \sqrt{2} \pi } s^2 \ue^{-2 b s^2}
      \left[\sqrt{\pi } \ue^{\frac{3 b s^2}{2}} \left(-b^2 s^4+6 b
          s^2-27\right) \left\{\text{erf}\left(\frac{\sqrt{b}
              s}{\sqrt{6}}\right) \right. \right. \nonumber \\
              &- \left . \left .\text{erf}\left(\frac{\sqrt{3} \sqrt{b} s}{\sqrt{2}} \right)
        \right\}  -\sqrt{6} \sqrt{b} s
          \left(\ue^{\frac{4 b s^2}{3}} \left(b s^2-9\right)+3 \left(7 b
              s^2+9\right)\right)\right], \label{eq:fnns_gue}\\
      P^{\text{GSE}}_{\text{\FN}} (s) &=  \frac{c^{5/2}}{2880 \sqrt{2} \pi } s^4 \ue^{-2 c s^2}
        \left[\sqrt{\pi } \ue^{\frac{3 c s^2}{2}} \left(c^4 s^8-12 c^3
            s^6+162 c^2 s^4-1620 c s^2+8505\right)
          \right. \nonumber\\
          &\left. \quad \quad \quad \quad \quad \quad \quad
            \quad \times
            \left\{\text{erf}\left(\sqrt{\frac{3}{2}}\sqrt{c} s\right)
            - \text{erf}\left(\frac{\sqrt{c}
                s}{\sqrt{6}}\right)\right\} \right. \nonumber\\
          & \left. \quad \quad \quad \quad \quad \quad \quad
            \quad- \sqrt{6} \sqrt{c} s \left(\ue^{\frac{4 c s^2}{3}}
\left(c^3
              s^6-15 c^2 s^4+225 c s^2-2835\right)\right. \right. \nonumber\\
        &\left. \left. \quad \quad \quad \quad \quad \quad \quad
            \quad +15 \left(91 c^3
              s^6+243 c^2 s^4+459 c s^2+567\right)\right)\right] .
      \label{eq:fnns_gse}
\end{align}
\end{subequations}

\item The limit of small spacings gives
  \begin{subequations} \label{eq:fnns0}
\begin{align}
	[P_{\text{\FN}}^\text{GOE}(s)]_{s\rightarrow 0} &=
	\frac{10\, a^2}{\pi}\, s^4+ O\left(s^6\right),   \label{eq:fnns0_goe}\\
	[P_{\text{\FN}}^\text{GUE}(s)]_{s\rightarrow 0} &=
\frac{124\, b^4}{15 \sqrt{3} \pi}\, \,  s^7 +O\left(s^9 \right), \label{eq:fnns0_gue}\\
      [P_{\text{\FN}}^\text{GSE}(s)]_{s\rightarrow 0} &=
\frac{ 5884\, c^7 }{14175 \sqrt{3} \pi } \, s^{13} +O(s^{15}). \label{eq:fnns0_gse}
\end{align}
\end{subequations}
\item For large spacings the behaviour is
  \begin{subequations} \label{eq:fnnsinfty}
\begin{align}
	[P_{\text{\FN}}^\text{GOE}(s)]_{s\rightarrow \infty} &\sim
	8 a\, \exp(-2a s^2/3),   \label{eq:fnnsinfty_goe}\\
	[P_{\text{\FN}}^\text{GUE}(s)]_{s\rightarrow \infty} &\sim
\frac{18 \sqrt{3}\, b}{\pi}\, \,  s^2\, \exp(-2b s^2/3), \label{eq:fnnsinfty_gue}\\
      [P_{\text{\FN}}^\text{GSE}(s)]_{s\rightarrow \infty} &\sim
\frac{ 54\sqrt{3} \, c^2 }{5 \pi }\, s^3 \exp(-2c s^2/3). \label{eq:fnnsinfty_gse}
\end{align}
\end{subequations}
\end{enumerate}
Thus the small $s$ behaviour is $s^{3\beta+1}$, while for large
$s$ the density is $\sim s^{\beta-1} \exp[-\beta s^2/(3 \mu_{\beta}^2)]$.

In contrast to the high degrees of level repulsion in the Gaussian ensembles, for a Poissonian spectrum it follows using the standard method for finding the probability of the maximum of two random variables, that
\begin{equation} \label{eq:FnN-Poisson}
  P_{\text{\FN}}^{\text{Poisson}}(s) = 2 \ue^{-s}(1- \ue^{-s}).
\end{equation}
This goes as $\sim 2s$ for small spacings, and $\sim 2 \exp(-s)$ for
large $s$, reflecting the larger probabilities of finding both small
and large values of $\text{FN}$ in comparison to the Gaussian
ensembles. The $\text{FN}$ behaviour may also be contrasted with the
usual behaviour of the consecutive level spacings of the Gaussian
ensembles, that is $\sim s^{\beta}$ for small $s$ and $s^{\beta}
\exp(-c_{\beta} s^2)$ for large $s$, again indicating the smaller
relative variability of the $\text{FN}$.

\subsection{Closest neighbour (\CN) spacings}
\label{sec:cn}

Following a similar approach, the closest neighbour spacing
probability densities can be worked out by first finding the cumulative
distribution function for $s_i^{\CN}$ as $Q[\min(s_1,
s_2)<0]$ or alternatively using
\begin{equation}
\text{Prob}(\text{min}\{s_1,s_2\}>s)=\int_s^{\infty} \ud s' P_{\text{CN}}(s')=\int_s^{\infty} \ud s_1 \int_s^{\infty} \ud s_2 P(s_1,s_2),
\end{equation}
which implies
\begin{equation}
P_{\text{CN}}(s)=2 \int_s^{\infty} \ud s_1 P(s_1,s),
\end{equation}
where again the symmetry of  $P(s_1,s_2)$ in Eq.~(\ref{eq:jpdfs1s2}) is invoked. From Eq.~(\ref{eq:fndist}) for the $\text{FN}$ density, it follows that
\begin{equation}
\frac{1}{2}P_{\FN}(s)+ \frac{1}{2}P_{\CN}(s)= \int_0^{\infty} \ud s_1 P(s_1,s)=  p(s),
\label{eq:mixture}
\end{equation}
where $p(s)$ is the consecutive level spacing probability density. Thus it is an equal mixture of those of the $\CN$ and $\FN$.

The probability density functions of $\CN$ for the three Gaussian ensembles are
\begin{subequations} \label{eq:cnn}
\begin{align}
  P_{\text{\CN}}^\text{GOE}(s) &=
	\frac{a}{\pi } s \ue^{-2 a s^2} \left[3 \sqrt{6 \pi } \sqrt{a} s-\pi
      \ue^{\frac{3 a s^2}{2}} \left(a s^2-3\right)
      \text{erfc}\left(\sqrt{\frac{3}{2}} \sqrt{a} s\right)\right],
      \label{eq:cnn_goe}\\
  P_{\text{\CN}}^\text{GUE}(s) &=
	 \frac{b^{3/2}}{4 \sqrt{2} \pi } s^2 \ue^{-2 b s^2} \left[\sqrt{\pi }
      \ue^{\frac{3 b s^2}{2}} \left(b^2 s^4-6 b s^2+27\right)
      \text{erfc}\left(\sqrt{\frac{3}{2}} \sqrt{b} s\right)\right. \nonumber \\
    & \left. \quad \quad \quad \quad \quad \quad \quad +3 \sqrt{6}
      \sqrt{b} s \left(7 b s^2+9\right)\right],   \label{eq:cnn_gue}\\
  P_{\text{\CN}}^\text{GSE}(s) &=
	  \frac{c^{5/2}}{2880 \sqrt{2} \pi } s^4 \ue^{-2 c s^2} \left[15
      \sqrt{6} \sqrt{c} s \left(91 c^3 s^6+243 c^2 s^4+459
        cs^2+567\right)\right. \nonumber\\
      & \left. \quad \quad\quad\quad\quad\quad +\sqrt{\pi } \ue^{\frac{3 c
            s^2}{2}} \left(c^4 s^8-12 c^3 s^6+162 c^2 s^4-1620 c
          s^2+8505\right)\right. \nonumber\\
      &\left. \quad\quad\quad\quad\quad\quad\quad\quad\quad \times
        \text{erfc}\left(\sqrt{\frac{3}{2}} \sqrt{c} s\right)\right].
        \label{eq:cnn_gse}
\end{align}
\end{subequations}
For small spacings, the probability densities show level-repulsion,
similar to the consecutive level spacing probability densities which
goes as $\sim s^{\beta}$, i.e.
  \begin{subequations} \label{eq:cnns0}
\begin{align}
	[P_{\text{\CN}}^\text{GOE}(s)]_{s\rightarrow 0} &=
	3 a s-\frac{5 a^2 s^3}{2} + \cdots,  \label{eq:cnns0_goe}\\
	[P_{\text{\CN}}^\text{GUE}(s)]_{s\rightarrow 0} &=
\frac{27 b^{3/2} s^2}{4 \sqrt{2 \pi }}-\frac{39 b^{5/2} s^4}{8
        \sqrt{2 \pi }} +\cdots,  \label{eq:cnns0_gue}\\
      [P_{\text{\CN}}^\text{GSE}(s)]_{s\rightarrow 0} &=
 \frac{189 c^{5/2} s^4}{64 \sqrt{2 \pi }}-\frac{261 c^{7/2}
        s^6}{128 \sqrt{2 \pi }} +\cdots. \label{eq:cnns0_gse}
\end{align}
\end{subequations}

A similar calculation yields $P_{\text{\CN}}(s)$ for a Poissonian
spectrum. In general,
from the inclusion--exclusion principle it follows that
the cumulative distribution function for $Z \equiv \min(X,Y)$ is related to the
cumulative distribution functions of $X$, $Y$
and the joint cumulative distribution of $X$, $Y$ as,
\begin{align}
    Q_Z(z) = Q_X(z) + Q_Y(z) - Q_{X,Y}(z,z) .
\end{align}
For the Poisson spectrum
$Q_{s_1}(s)=1-\ue^{-s}$, $Q_{s_2}(s)=1-\ue^{-s}$ and
$Q_{s_1,s_2}(s,s) = Q_{s_1}(s)Q_{s_2}(s) = (1-\ue^{-s})^2$
and therefore $Q_Z(s)=(1-\ue^{-2s})$.
Taking the derivative gives
\begin{equation} \label{eq:CnN-Poisson}
  P_{\text{\CN}}^{\text{Poisson}}(s) = 2\ue^{-2s},
\end{equation}
consistent with the fact that an equal mixture with the $\FN$ density in Eq.~(\ref{eq:FnN-Poisson})
yields $\ue^{-s}$, the first consecutive level spacing density of the
Poisson distribution. The result for the Poissonian $\CN$  has been
used in the
context of perturbation theory for the entanglement
in bi-partite systems \cite{LakSriKetBaeTom2016,TomLakSriBae2018}.

\begin{figure}[t]
  \begin{center}
  \includegraphics[scale=1.15]{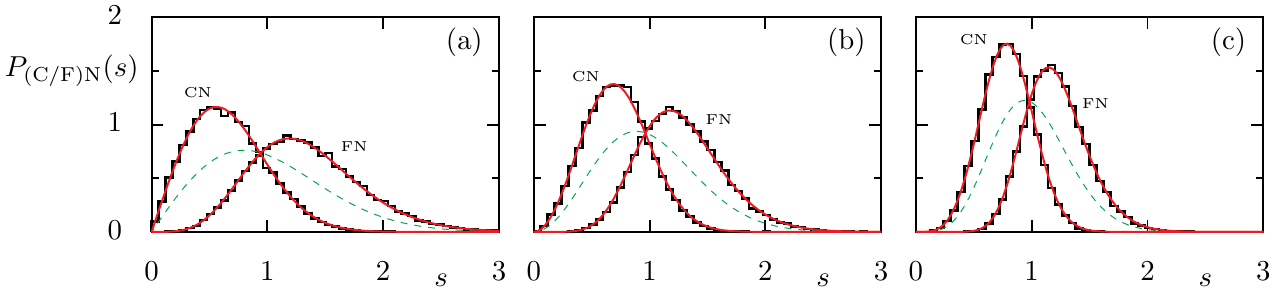}
  \end{center}
  \caption{\label{fig:RMT}
$P_{\text{(C/F)N}}(s)$.
The \CNtxt{} spacing density $P_{\text{CN}}(s)$
and \FNtxt{} spacing density $P_{\text{FN}}(s)$
for COE, CUE, and CSE matrices in comparison
 with the corresponding
predictions, Eqs.~\eqref{eq:fnns} and \eqref{eq:cnn}.
For the COE and CUE ensembles, 600 matrices of dimension $100$ are created
and for the CSE, 600 matrices of dimension $200$.
For comparison the corresponding
consecutive level spacing densities are shown as dashed curves.
}
\end{figure}

Figure \ref{fig:RMT} shows a comparison of the analytical results
for the \CNtxt{} densities of Eq.~\eqref{eq:cnn}
and \FNtxt{} densities of Eq.~\eqref{eq:fnns}
with calculations using
matrices of the Circular orthogonal ensemble (COE),
Circular unitary ensemble (CUE) and
Circular symplectic ensemble (CSE);
they are generated as described in Ref.~\cite{Mez2007}.
Despite approximating the full JPD by a $3\times 3$ matrix, the
analytical results describe the numerical results very well.  This is
very similar to how the
Wigner surmise captures the consecutive level spacing
probability density of the infinite dimensional ensembles from a $2
\times 2$ matrix ensemble.

The mean values of the \CNtxt{} and the \FNtxt{} spacings for the Poissonian, GOE, GUE, and
GSE ensembles are summarized in Table~\ref{tab:mean-values}.  Notice that
mean values of the two cases are approaching each other with
increasing amount of level repulsion in the spectrum, and their sum is always $2$, as required
by Eq.~(\ref{eq:mixture}).

\newcommand\T{\rule{0pt}{3.0ex}}       
\newcommand\B{\rule[-1.6ex]{0pt}{0pt}} 
\begin{table}
\begin{center}
\begin{tabular}{|c||c|c||c|}
\hline
Spectrum & $\langle s_{\text{\CN}} \rangle$ & $\langle s_{\text{\FN}}
\rangle$ &
 Difference \T\B\\ \hline\hline
Poissonian & $\frac{1}{2}$ & $\frac{3}{2}$ & $1$  \T\B\\ \hline
GOE & $\frac{2}{3}$ & $\frac{4}{3}$        & $\frac{2}{3}$  \T\B\\ \hline
GUE & $ 4- \frac{17}{3\sqrt{3}}$ & $ \frac{17}{3\sqrt{3}}-2$
   & $\frac{34}{3\sqrt{3}} - 6 \approx 0.5433$  \T\B\\
\hline
GSE & $4-\frac{1801}{324 \sqrt{3}}$ & $\frac{1801}{324 \sqrt{3}}-2$
    & $ \frac{1801}{162 \sqrt{3}} -6  \approx 0.4186$ \T\B\\
\hline
\end{tabular}
\end{center}
\caption{Mean values of the \CNtxt{} and \FNtxt{} spacings for Possonian, GOE, GUE, and GSE ensembles.}
\label{tab:mean-values}
\end{table}

\section{Application to various systems}
\label{sec:applications}

The \CNtxt{} and \FNtxt{} densities are examined here for a series of dynamical systems and the Riemann zeta function.
For integrable systems the probability density $P(s)$ of consecutive level spacings
is given by~\cite{BerTab1977}
\begin{equation} \label{P(S)-POIS}
  P_{\text{Poisson}}(s) = \ue^{-s}.
\end{equation}
In this case the density does not vanish at the origin, i.e.~$P(s)\to 1$ for $s\to 0$.  In contrast, the
densities of classically strongly chaotic systems follow the corresponding
RMT densities \cite{BohGiaSch1984}.
The GOE result for the consecutive level spacing density
is to a very good approximation given by
\begin{equation} \label{P(S)-GOE}
  P_{\text{GOE}}(s)
    =  \frac{\pi}{2} s \exp\left(-\frac{\pi}{4}s^2\right).
\end{equation}
For the GUE one has to a very good approximation
\begin{equation} \label{P(S)-GUE}
  P_{\text{GUE}}(s)
  =   \frac{32}{\pi^2} s^2
                      \exp\left(-\frac{4}{\pi}s^2\right).
\end{equation}
For these two cases and the GSE the spectrum displays level repulsion
as $P(s) \sim s^\beta$ for small $s$.

\subsection{Quantum billiards}

A classical billiard system is given by the free motion
of a point particle inside some domain $\Omega$
with specular reflections at the boundary, i.e.\
the angle of incidence equals the angle of reflection.
Depending on the shape of the boundary, the dynamics
can range from integrable
such as for the circular, rectangular or elliptical billiards,
to fully chaotic motion as for the Sinai-biliard \cite{Sin1970},
Bunimovich stadium billiard \cite{Bun1979},
or the cardioid billiard \cite{Rob1983,Woj1986,Mar1988,BaeDul1997}.

A corresponding quantum billiard is described
by the stationary Schr\"odinger equation (in units $\hbar=2m=1$)
\begin{equation}
\label{Schroedinger}
  -\Delta \psi_n({\bf q}) = E_n  \psi_n({\bf q})\;\;, \quad \bfq\in \Omega,
\end{equation}
with Dirichlet boundary conditions,
i.e.\ $\psi_n({\bf q})=0$ for $\bfq\in \partial\Omega$.
The unfolded spectrum is obtained by the mapping
\begin{equation}
  x_n \;  :=  \;\overline{N}(E_n),
\end{equation}
where $\overline{N}(E)$ is the mean behaviour of the spectral
staircase function and is given by
the generalized Weyl formula \cite{BalHil1976}
\begin{equation} \label{Weyl}
  \overline{N}(E) =  \frac{\CA}{4\pi} E
                      - \frac{\CL}{4\pi} \sqrt{E}
                      + \CC,
\end{equation}
where $\CA$ denotes the area of the billiard, and
$\CL:=\CL^- - \CL^+$, where $\CL^-$ and $\CL^+$
are the lengths of the pieces of the boundary $\partial \Omega$
with Dirichlet and Neumann boundary conditions,
respectively.
The constant $\CC$ takes curvature and corner corrections into account.

\begin{figure}[t]
\includegraphics[scale=1.2]{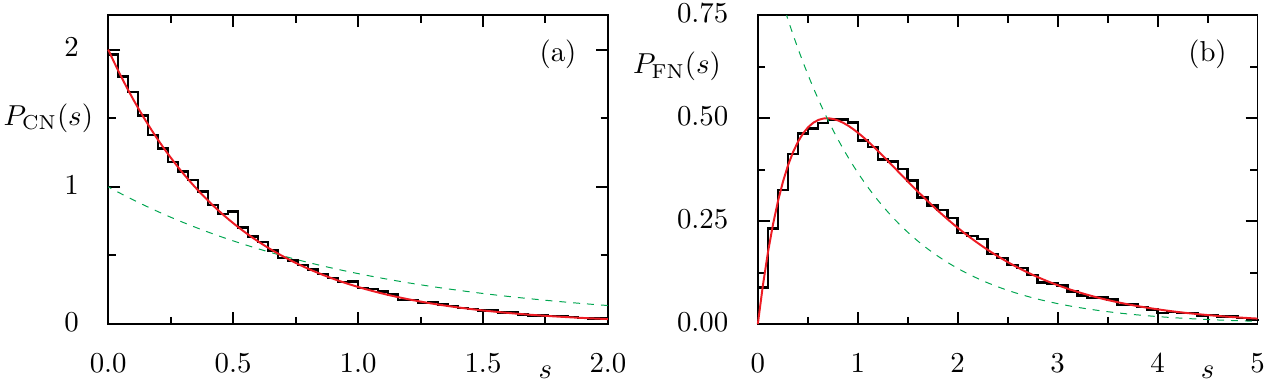}
\caption{Circle billiard: (a) \CNtxt{} spacing density $P_{\CN}(s)$
         and (b) \FNtxt{} spacing denisty $P_{\FN}(s)$
         in comparison with
         Eqs.~\eqref{eq:CnN-Poisson} and \eqref{eq:FnN-Poisson},
         respectively,
         for a Poissonian spectrum (red line).
         Also shown is the exponential, Eq.~\eqref{P(S)-POIS},
         for the consecutive level spacing probability density (dashed line).
         }
         \label{fig:CN-FN-circle}
\end{figure}

As an example integrable system consider the circular billiard
with unit radius and Dirichlet boundary conditions.
For the desymmetrized (half circular) billiard
and Neumann boundary conditions on the symmetry line,
the eigenvalues are $E_{kl} = j_{kl}^2$,
where $j_{kl}$ is the $l$-th zero of the Bessel function $J_k(x)$
with $k=0, 1, 2, ...$ and $l=1, 2, ...$.
The corresponding constants in Eq.~\eqref{Weyl} are given  by
$\CA = \frac{\pi}{2}$, $\CL=\pi-2$ and $\CC=-\frac{1}{24}$.
The first $10^5$ eigenvalues give
for the \CNtxt{} spacing density $P_{\CN}(s)$
and \FNtxt{} spacing density $P_{\FN}(s)$
the results shown in Fig.~\ref{fig:CN-FN-circle}.
Good agreement is found with the predictions of Eqs.~\eqref{eq:CnN-Poisson}
and \eqref{eq:FnN-Poisson}, respectively.

\begin{figure}[t]
\includegraphics[scale=1.2]{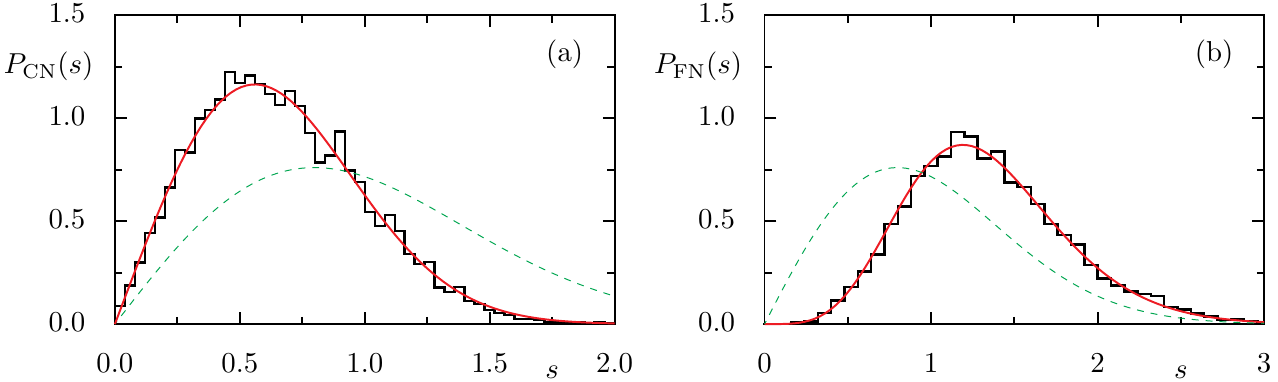}
\caption{Cardioid billiard: (a) \CNtxt{} spacing density $P_{\CN}(s)$
         and (b) \FNtxt{} spacing density $P_{\FN}(s)$
         in comparison with the predictions
         Eqs.~\eqref{eq:cnn_goe} and Eq.~\eqref{eq:fnns_goe}, respectively,
         for the GOE (red line).
         Also shown is the approximation Eq.~\eqref{P(S)-GOE}
         for the consecutive level spacing density of the GOE (dashed line).
         }
         \label{fig:CN-FN-cardioid}
\end{figure}

Next consider the chaotic cardioid billiard, which is
the limiting case of a billiard family introduced in Ref.~\cite{Rob1983}.
Its boundary is given in polar coordinates $(\rho, \phi)$ by
\begin{equation}
\label{eq:cardioid-boundary}
        \rho(\phi)=1+ \cos\phi \;\; , \qquad \phi \in [-\pi,\pi] \;\;.
\end{equation}
The cardioid billiard is proven to be strongly chaotic
\cite{Woj1986,Sza1992,Mar1993, LivWoj1995,CheHas1996}.
For the desymmetrized cardioid billiard with Dirichlet boundary
conditions, the statistics is based on the first
11,000 eigenvalues \cite{PrivComProRob},
determined using the conformal mapping technique \cite{Rob1984,ProRob1993a}.
For a detailed investigation of the spectral statistics
see \cite{BaeSteSti1995}.
For the mean behaviour $\overline{N}(E)$, Eq.~\eqref{Weyl},
of the spectral staircase function the constants are
$\CA = 3\pi/4$, $\CL=6$, and $\CC=3/16$.
Figure~\ref{fig:CN-FN-cardioid} shows
the \CNtxt{} spacing density $P_{\CN}(s)$
and \FNtxt{} spacing density $P_{\FN}(s)$
and there is good agreement with the predictions of Eqs.~\eqref{eq:cnn_goe}
and \eqref{eq:fnns_goe}, respectively.
The comparison with the consecutive level spacing density $P_{\text{GOE}}(s)$
shows that $P_{\CN}(s)$ has its maximum for smaller $s$ and
$P_{\FN}(s)$ has its maximum for larger $s$
and the mean values, as shown in Table~\ref{tab:mean-values},
are also below and above 1.

\subsection{Quantum maps}

Another extensively studied class of systems are quantum maps which are unitary
operators on a finite--dimensional Hilbert space.
Their classical limit is given by a discrete-time mapping on phase space.
A prototypical example is the so-called kicked-rotor \cite{Chi1979}.
It is obtained from the Hamiltonian
\begin{equation}
  H(q, p, t) = \frac{p^2}{2} + V(q) \sum_t \delta(t-n),
\end{equation}
where the sum describes a periodic sequence of kicks with unit time as
the kicking period and $V(q) = K \cos(2 \pi q)/4 \pi^2$.
Considering the dynamics immediately prior to consecutive kicks
leads to the standard map,
\begin{equation}
\label{eq:clmap}
(q, p) \mapsto
(q', p') =\left(q+p', p + \tfrac{K}{2\pi} \sin(2\pi q) \right),
\end{equation}
which is a two-dimensional area-preserving map.
The map is considered on a two-dimensional torus
with periodic boundary conditions in both $q$ and $p$.
If the kicking strength $K$
is sufficiently large, the map is strongly chaotic with a Lyapunov
exponent of $\approx \ln (K/2)$~\cite{Chi1979}.
This excludes some special values corresponding to small
scale islands formed by the so-called accelerator modes \cite{Chi1979}.
This is supported by recent rigorous results
showing that the stochastic sea of the standard
map has full Hausdorff dimension for
sufficiently large generic parameters \cite{Gor2012}.
As example we use $K=10$
for which the dynamics of the map numerically looks chaotic,
i.e.\ no regular islands are observed on any relevant scales.

The quantum mechanics on a torus phase space
leads to a finite Hilbert space of dimension $N$,
see e.g.\ Refs.~\cite{BerBalTabVor1979, HanBer1980, ChaShi1986, KeaMezRob1999,
  DegGra2003bwcrossref}.
The effective Planck constant is $h=1/N$.
The quantum map $U$ is an unitary operator,
\begin{align}
\label{2dqmap}
U(n^{\prime},n) = &  \frac{1}{N}\,  \exp\left(- \ui N
\frac{K}{2 \pi} \cos\left(\frac{2\pi}{N}(n+\alpha)\right)\right) \nonumber \\
& \times
\, \sum_{m=0}^{N-1}\exp\left(-\frac{\pi \ui }{N} (m+\beta)^2\right) \,
\exp\left(\frac{2 \pi \ui}{N} (m+\beta)(n-n^{\prime})\right),
\end{align}
where $n,n'\in\{0,1, ..., N-1\}$.
The parameters
$\beta$ and $\alpha$ are the quantum phases due to periodicity in
position and momentum respectively.
The cases when these are $0$ or $1/2$ correspond to
periodic and anti-periodic boundary conditions respectively.
We choose $(\alpha, \beta) = (0.2, 0.3)$
ensuring that parity and time reversal symmetry are both broken.
Thus, the spectral statistics should follow
those of the circular unitary ensemble (same as GUE).

\begin{figure}[t]
\includegraphics[scale=1.2]{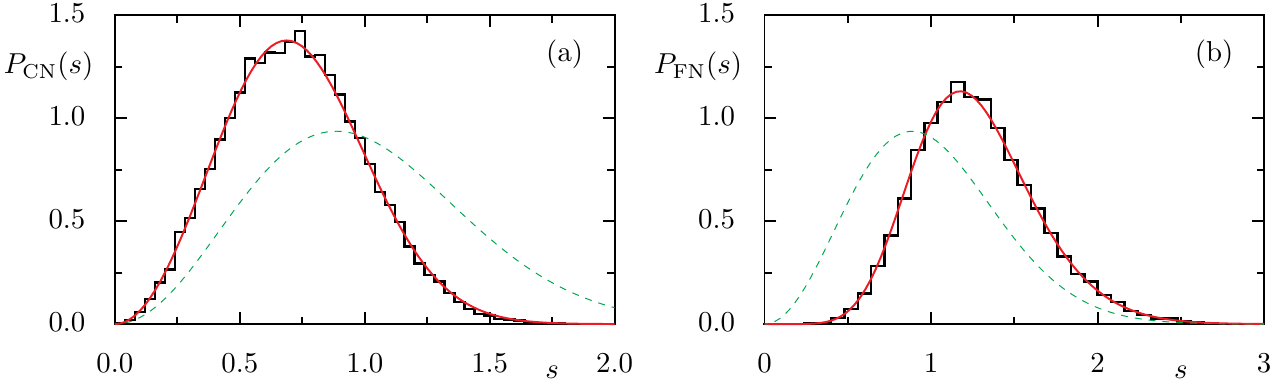}
\caption{Quantized standard map:
         (a) \CNtxt{} spacing density $P_{\CN}(s)$  and (b) \FNtxt{}
         spacing density $P_{\FN}(s)$
         in comparison with the predictions
         Eqs.~\eqref{eq:cnn_gue} and \eqref{eq:fnns_gue}, respectively,
         for the GUE (red line).
         Also shown is the approximation, Eq.~\eqref{P(S)-GUE}
         for the consecutive level spacing density of the GUE
          (dashed line).
         }
         \label{fig:CN-FN-std-map}
\end{figure}

Figure~\ref{fig:CN-FN-std-map} shows the
\CNtxt{} spacing density $P_{\CN}(s)$
and the \FNtxt{} spacing density $P_{\FN}(s)$
for the quantized standard map for $N=30,000$.
The agreement with the expected GUE results,
Eqs.~\eqref{eq:cnn_gue} and \eqref{eq:fnns_gue}, is very good.

\subsection{Riemann zeta function}

\begin{figure}
\includegraphics[scale=1.2]{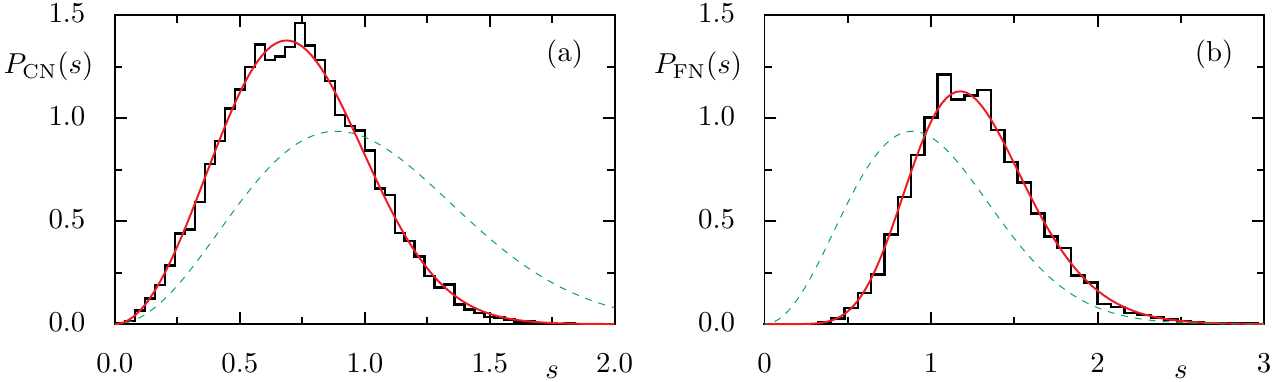}
\caption{Riemann zeros: (a) \CNtxt{}  spacing density $P_{\CN}(s)$ and
         (b) \FNtxt{} spacing density $P_{\FN}(s)$
         in comparison with the GUE predictions (red line)
         Eqs.~\eqref{eq:cnn_gue} and \eqref{eq:fnns_gue}, respectively.
         Also shown is the approximation, Eq.~\eqref{P(S)-GUE}
         for the consecutive level spacing density of the GUE
         (dashed line).
         }
         \label{fig:CN-FN-riemann}
\end{figure}

An interesting connection between the spectral
properties of chaotic dynamical systems and
number theory arises with the statistics of the zeros of
the Riemann zeta function.
According to a famous conjecture, its zeros
are expected to lie on the critical line \cite{Rie1859}, i.e.\
\begin{equation}
  \zeta(1/2 + \ui t_n) = 0 \;\;.
\end{equation}
Their counting function is given by
the Riemann--von Mangoldt function \cite{Rie1859,Man1905,TitHea1988}
\begin{equation} \label{eq:riemann-von-mangoldt}
 N(T) = \frac{T}{2\pi} \ln \left( \frac{T}{2\pi} \right) - \frac{T}{2\pi}
        + \frac{7}{8}.
\end{equation}
As is well known for the spacing densities,
the convergence to the CUE results is rather slow \cite{Odl1987}.
Therefore, the $10,000$ zeros
starting from the $(10^{22} + 1)$-th \cite{OdlTables} are used.
To numerically perform the unfolding for the tabulated zeros, which
have an offset $\delta = 1370919909931995300000$,
this has to be accounted for when applying \eqref{eq:riemann-von-mangoldt}.
Assuming that $T = x + \delta$ gives
\begin{equation}
  N(x+\delta) \approx \frac{x^2}{2\pi \delta} +
         \frac{x}{2\pi} \ln \left( \frac{\delta}{2\pi} \right)
         + \text{const}(\delta) \;\;,
\end{equation}
where the approximation $\ln(1 + x/\delta) \approx x/\delta$
has been used.  After this unfolding the mean spacing is $1.00011$.
Figure~\ref{fig:CN-FN-riemann} shows the
\CNtxt{} $P_{\CN}(s)$ and the
\FNtxt{} spacing densities $P_{\FN}(s)$
and very good agreement with the expected GUE results,
Eq.~\eqref{eq:cnn_gue} and Eq.~\eqref{eq:fnns_gue}, respectively, is found.

\section{$k$-th closest neighbour (\kCN) spacings}
\label{sec:kcn}

By means of the \kCNtxt{} spacing density
longer-range statistical properties can be investigated.

\subsection{Poisson sequences}

For the case of a Poissonian spectrum
it is possible to obtain an analytical prediction.
The steps of the proof are illustrated for the \SCN{}:
\begin{itemize}
\item The unfolded eigenvalues form a unit mean Poisson process such
that the consecutive level spacings $s_i$ are
distributed according to $\exp(-s)$.

\item Each semi-infinite sequence obtained by splitting
this infinite sequence of levels from a fixed level is
also described by a unit mean Poisson process.

\item Flipping one of these two semi-infinite sequences and merging
 results in a semi-infinite sequence described by a Poisson
process with rescaled spacing, $1/2$.

\item Therefore, the consecutive level spacings
 for this new semi-infinite sequence would be the same as the original infinite sequence, except for the change of the mean spacing.  Hence, for example $\exp (-s)\to 2\exp(-2s)$.  Explicitly,
\begin{equation}
\label{eq:CknN-Poisson}
  P_{\text{kCN}}^{\text{Poisson}} = \frac{2^k}{(k-1)!} s^{k-1} \exp(-2 s) .
\end{equation}
\item[] The means are given by
\begin{equation} \label{eq:mean-Poisson}
   \mu_k^{\text{Poisson}} = \langle s^{\kCN}  \rangle = k/2
\end{equation}
and the variances by
\begin{equation} \label{eq:variance-Poisson}
    \sigma^{2,{\text{Poisson}}}_k
         = \left\langle \left[s^{\kCN} - k/2\right]^2  \right\rangle
         = k/4 .
\end{equation}
\end{itemize}

\begin{figure}
\includegraphics[scale=1.2]{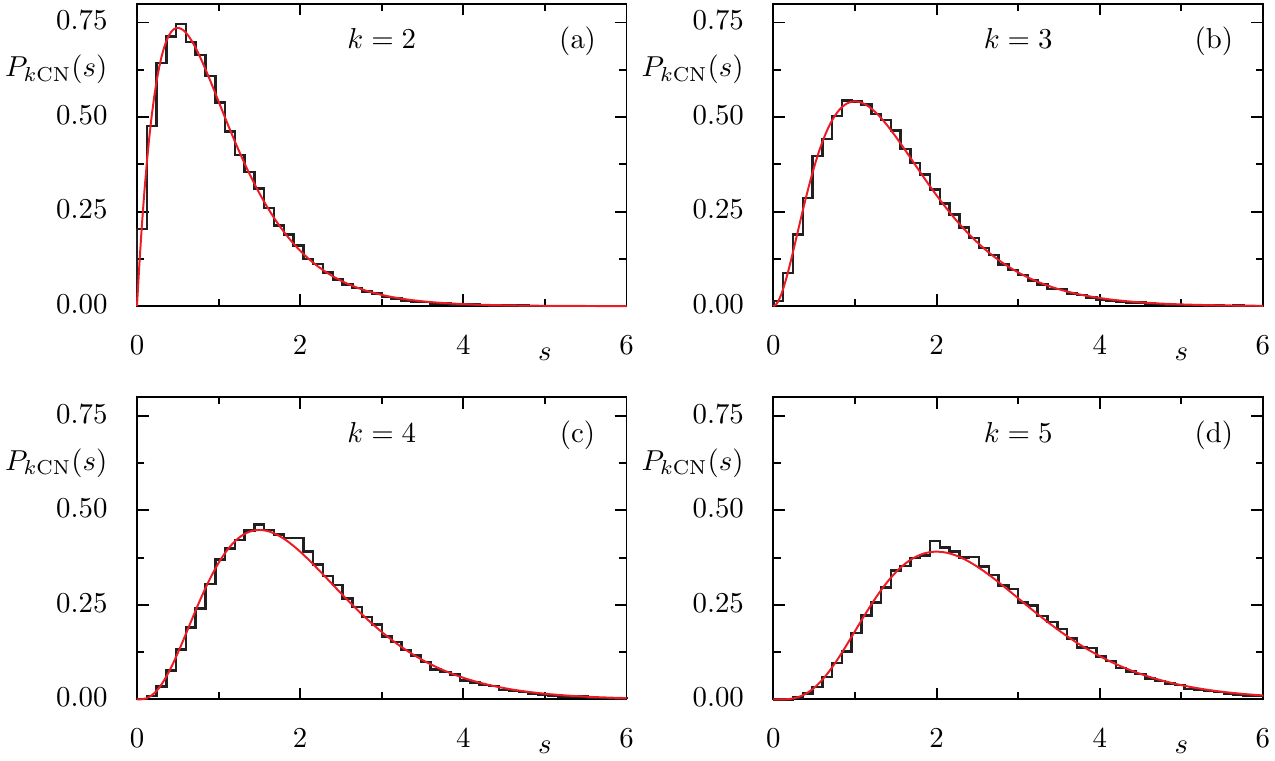}
\caption{Probability densities $P_{\kCN}(s)$ of the \kCNtxt{} spacings
         for the integrable circle billiard $k=2, 3, 4, 5$
         in comparison with expected results
         Eq.~\eqref{eq:CknN-Poisson} for a Poissonian spectrum (red line).
         }
         \label{fig:kCN-circle}
\end{figure}

Figure~\ref{fig:kCN-circle} shows $P_{\kCN}(s)$ for $k=2, 3, 4, 5$ for the
integrable circle billiard in comparison with Eq.~\eqref{eq:CknN-Poisson}.
\begin{figure}
\includegraphics[scale=1.2]{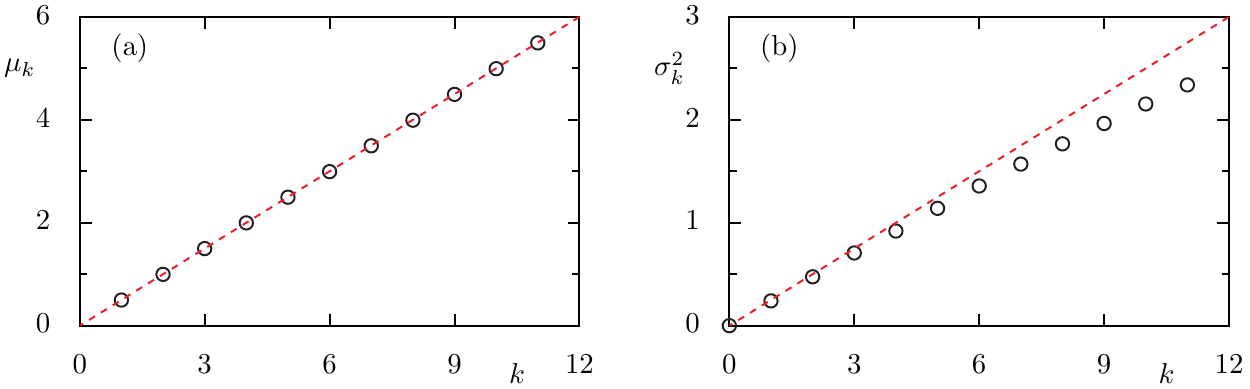}
\caption{Means and variances $\sigma^2_k$, of the \kCNtxt{} spacings
         for the integrable circle billiard (symbols)
         in comparison with the analytic Poisson results,
         Eqs.~\eqref{eq:mean-Poisson}
         and \eqref{eq:variance-Poisson}, respectively
         (dashed red lines).
         }
         \label{fig:mean-variance-circle}
\end{figure}
Overall, very good agreement is found, but with increasing $k$ some very
small deviations begin to show.
They are a manifestation of the increasing deviations of the
variance for larger $k$ from $k/4$; see Fig.~\ref{fig:mean-variance-circle}.
This is similar to the well-known saturation of the
spectral rigidity and number variance \cite{Ber1985,Ber1988,AurSte1990},
which also determines the variance of the
$k$-th consecutive level spacing statistics \cite{BroFloFreMelPanWon1981}.

\subsection{RMT sequences}

\begin{figure}[b]
\includegraphics[scale=1.2]{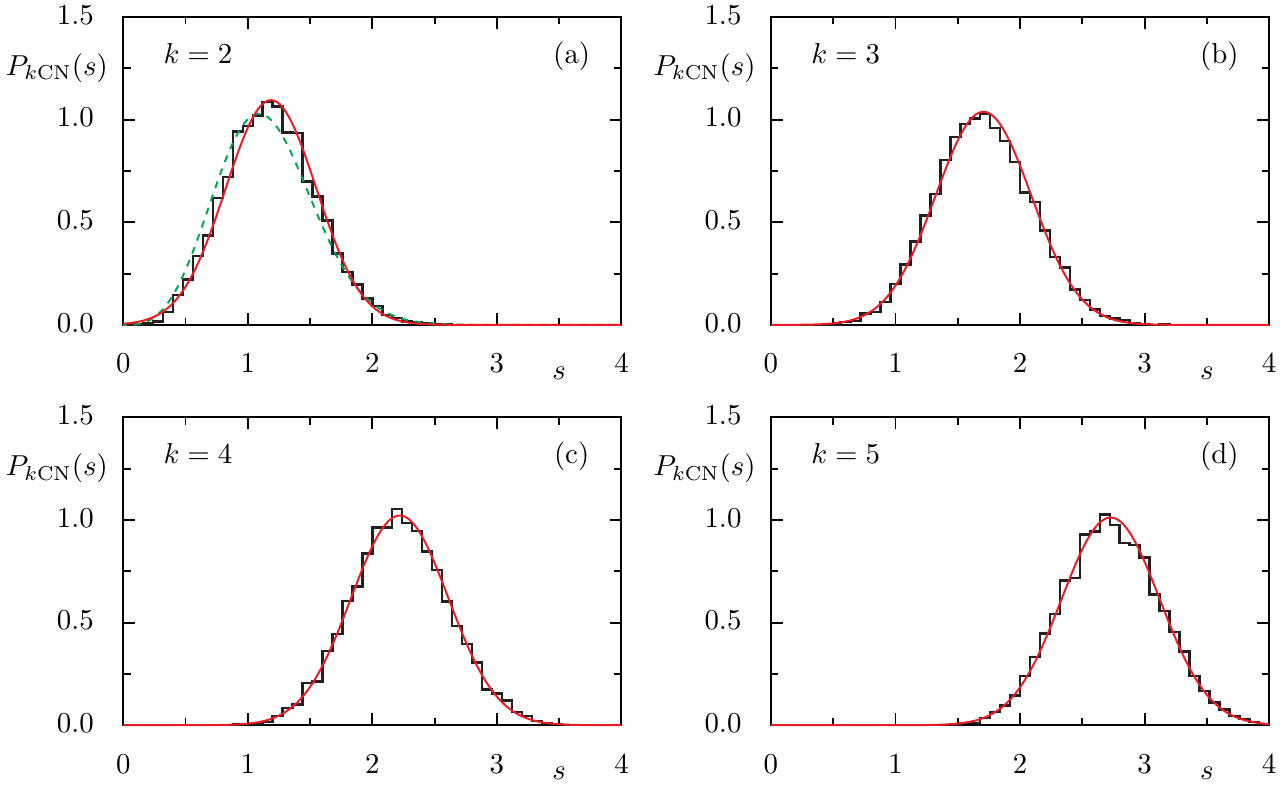}
\caption{Densities $P_{\kCN}(s)$ of the \kCNtxt{} spacings
         for the cardioid billiard in comparison with the
         Gaussian densities, Eq.~\eqref{eq:CknN-Poisson}, red line,
         with mean and variance determined from the data.
         For $k=2$ the density, Eq.~\eqref{eq:scn-goe},
         is shown as green dashed line.
         }
         \label{fig:kCN-cardioid}
\end{figure}

Figure~\ref{fig:kCN-cardioid} shows the \kCNtxt{} spacing density
for the cardioid billiard.  From its definition, it is clear that the likelihood of the distance to the
$k$-th neighbour is the sum of correlated, yet randomly behaving
consecutive level spacings.
Supposing a central limit theorem for increasing $k$,
the density is expected to become more and more Gaussian~\cite{BroFloFreMelPanWon1981}
\begin{equation}
  G(k, s) \; \simeq \;
           \frac{1}{\sqrt{2\pi \sigma^2_k}}
           \exp\left(-\frac{(s-\mu_k)^2}{2\sigma^2_k}
                \right).
\end{equation}
This is in very good agreement with the numerically obtained histograms.

\begin{figure}[b]
\includegraphics[scale=1.2]{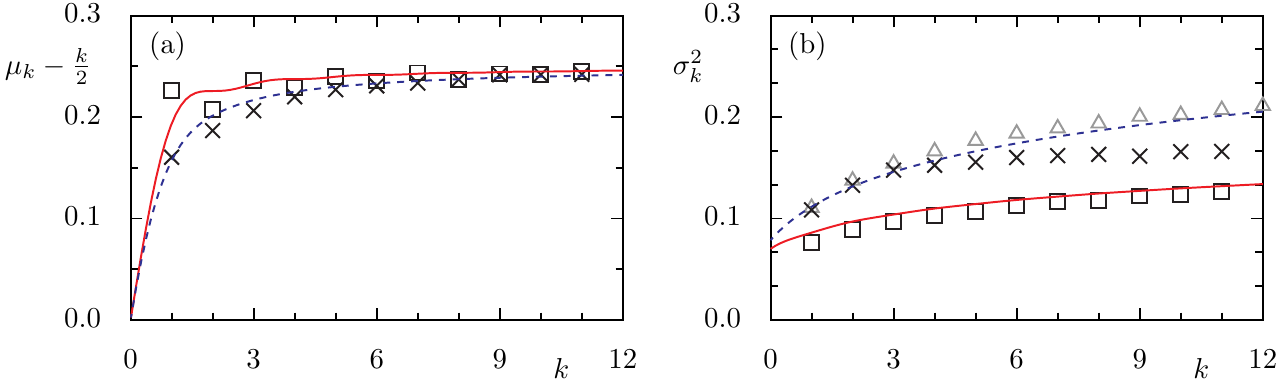}
\caption{Means and variances of the \kCNtxt{} spacings for the
standard map ($\square$) and cardioid billiard ($\times$).
The full red line, and blue dashed line are analytical
calculations
for GUE and GOE ensembles respectively.
In (a) $k/2$ is subtracted from the means in order to make the approach
to the expected offset $0.25$ and the odd-even effect more visible.
In (b) for comparison results obtained for 500
realizations of COE random matrices of size $100 \times 100$ are shown
($\triangle$).}
\label{fig:varmean}
\end{figure}

In addition, it is possible to give heuristic arguments for the means and variances of the \kCN.  Even though correlations exist in the spectra of chaotic systems, the definition of \kCN\ is equivalent to cutting an infinite sequence at a level and flipping one of the two semi-infinite subsequences just as
for the Poisson case.  The mean spacing reduces to $1/2$ as before.  However, the spectral correlations introduce some changes.  First, relative to the result for the Poisson sequence of Eq.~(\ref{eq:mean-Poisson}), there is a shift of the mean spacing.  RMT spectra exhibit a correlation hole.  Consider the case before superposing the flipped sequence.   Due to the symmetric, nearly evenly spaced sequence of $P_{\kCN}(s)$, the mean of $P_{\kCN}(s)$ is approximately equal to the width $s$ of the interval required to find $k$ spacings on average.  This gives implicitly
\begin{equation}
k = \int_{0}^{s} {\rm d}s^\prime R_2(s^\prime) = s - \int_0^{s} {\rm d}s^\prime Y_2(s^\prime).
\end{equation}
After superposing the left and right sequences, the density is doubled, and the mean is half of the
result.  Rearranging terms, this gives
\begin{equation}
s = \frac{1}{2}\left( k+ \int_0^s {\rm d}s^\prime
Y_2(s^\prime) \right).
\end{equation}
An approximate form of which follows by replacing the upper integration limit with the leading dependence of s ($\approx k/2$),
\begin{equation}
s = \mu_k \approx  \frac{k}{2} + \frac{1}{2}\int_0^{k/2} {\rm d}s^\prime Y_2(s^\prime).
\end{equation}
Using exact expressions of $Y_2(s)$ \cite[Eq.~(L1)]{BroFloFreMelPanWon1981} for
the respective ensembles,  we obtain
\begin{subequations} \label{eq:meankcn}
\begin{align}
	 \mu_k^\text{GOE} &= \frac{k}{2} + \frac{4 \pi  k \text{Si}(k \pi )+2 \left(\pi -2 \text{Si}\left(\frac{k \pi }{2}\right)\right) \sin \left(\frac{\pi  k}{2}\right)-\pi ^2 k+4 \cos (\pi  k)-4}{4 \pi ^2 k},\label{eq:meankcn_goe}\\
	 \mu_k^\text{GUE} &= \frac{k}{2} + \frac{\pi  k \text{Si}(k \pi )+\cos (\pi  k)-1}{2 \pi ^2 k}\label{eq:meankcn_gue}
\qquad \text{ where Si}(x) = \int_0^x \frac{\sin t}{t} \ud t .
\end{align}
\end{subequations}
These expressions are compared with results from the cardioid billiard and standard map,
respectively, in Fig.~\ref{fig:varmean}(a).
One sees that the trend towards $k/2 - 1/4$ is confirmed.  The
behaviour for small $k$ is fairly good, but shows some slight deviations. In particular for the
standard map, an odd-even effect is more pronounced than for the theoretical prediction.

The scale of the variances can be given by noting their similarity to the number variance.
 Neglecting three-point correlations and the sawtooth corrections~\cite{French78}, the
spacing variances should be roughly a quarter of the number variance of appropriate argument.
 Thus, they should grow no more than logarithmically with increasing
$k$.  For an illustration of these quantities and their comparison to these approximate forms,
see Fig.~\ref{fig:varmean}(b).
Considering the simplifying assumptions just invoked, for the quantized standard map the variance
$\sigma_k^2$ follows $\tfrac{1}{4}\Sigma^2_{\text{GUE}}((k+1)/2)$ up to a rather
small shift very well.
However, the agreement of $\sigma_k^2$ for the cardioid billiard with
$\tfrac{1}{4}\Sigma^2_{\text{GOE}}((k+1)/2)$ is not quite as good.
The deviations for larger $k$ correspond to the early
saturation of the number variance observed for this system
\cite{BaeSteSti1995,AurBaeSte1997}.
To emphasize the distinction, the figure also shows
results obtained for an ensemble of 500 COE matrices of
dimension 100, which again agrees very
well with the prediction up to a small shift.

To give an idea of the role of $3$-point correlations in the $P_{\kCN}(s)$,
the example of the \SCNtxt{} spacing density is calculated with the approximation that
the spacings themselves are \emph{independent}.  This incorporates most of the $2$-point correlation information, but completely removes $3$-point correlations.  By definition no more than two levels from either side of a chosen level can be the \SCNtxt{}, thus isolating 5 levels
$\{x_{i-2}, x_{i-1}, x_i, x_{i+1}, x_{i+2}\}$ of the spectrum; see Fig.~\ref{fig:sketch-x-i}.
The corresponding set $\{x_i - x_{i-2}, x_i - x_{i-1}, x_{i+1}-x_i, x_{i+2}-x_i \}$
of 4 spacings has two elements as the consecutive level spacings, namely
$x_i - x_{i-1} = s_{i-1}$ and $x_{i+1} - x_i = s_{i}$,
and two as sum of the consecutive level spacings, namely
$x_i - x_{i-2} = s_{i-2} + s_{i-1}$ and
$x_{i+2}-x_i = s_{i+1} + s_{i+2}$.
The \SCNtxt{} spacing density can be obtained by
  calculating the order statistics of the correlated spacings defined earlier.  To a
  first approximation, this sequence can be treated as being formed by independent,
  but not identical random variables.  Consecutive level spacings themselves are
  distributed according to the Wigner surmise, and sum of two are calculated by convolution.  These approximations give a form for the
  density and the distribution (cumulative density) of the elements of the sequence formed by
  consecutive level spacings and their sum.
Using the Bapat-Beg theorem \cite{BapBeg1989} for the density
of $k^\text{th}$ order statistics in terms of the permanent of a matrix
constructed using the probability densities and cumulative distribution functions
of the individual elements, the \SCN{} spacing density is,
\begin{equation} \label{eq:scn-goe}
\begin{aligned}
P_{\text{\SCN{}}}^\text{GOE}(s) &= \frac{1}{16} \pi  \ue^{-\pi  s^2}
\left[\pi  \ue^{\frac{\pi  s^2}{4}} s \left(-9 \pi  s^2+4 \ue^{\frac{\pi
s^2}{4}} \left(\pi  s^2-2\right)+12\right) \text{erf}\left(\frac{1}{2}
\sqrt{\frac{\pi }{2}} s\right)^2  \right.\\
& \left. \quad \quad+2 \sqrt{2} \ue^{\frac{\pi  s^2}{8}} \left(-18 \pi
s^2+\ue^{\frac{\pi  s^2}{4}} \left(13 \pi  s^2-12\right)+12\right)
\text{erf}\left(\frac{1}{2} \sqrt{\frac{\pi }{2}} s\right) \right. \\
& \left.\quad \quad +72 \left(\ue^{\frac{\pi  s^2}{4}}-1\right) s\right].
\end{aligned}
\end{equation}
These approximate densities in the limit of small and large $s$ behaves as,
\begin{equation}
  [P_{\text{\SCN{}}}^\text{GOE}(s)]_{s\to 0} = \frac{\pi^2}{4}s^3 +
  O\left(s^5\right) ,
  \qquad
  [P_{\text{\SCN{}}}^\text{GOE}(s)]_{s\to
    \infty} \sim \frac{\pi ^3}{4}  s^3 \ue^{-\frac{\pi}{2}   s^2}.
\end{equation}
The approximation, Eq.~\eqref{eq:scn-goe},
is also shown in Fig.~\ref{fig:kCN-cardioid}(a)
and some significant deviation is found.
Nevertheless, the approximation
reproduces the mean \SCN{} within $\sim 3\%$. Similar
calculations can be carried out for the GUE ensemble,
leading to a rather lengthy expression, which is omitted here.
The quality of agreement with the numerical results is similar
to the GOE case.

\section{Ratio statistics of \CN/\SCN{}}

To avoid the unfolding procedure, the ratio statistics of successive level spacings
has been introduced as \cite{OgaHus2007}
\begin{equation} \label{eq:ratio-statistics}
  r_i  = \frac{\text{min}(s_i, s_{i-1})}
              {\text{max}(s_i, s_{i-1})} ,
\end{equation}
which in terms of the ordered spacings is
the same as the ratio $s_i^{\CN}/s_i^{\FN}$.
Notice that this ratio mostly contains information of up to $3$-point correlations.  It has become very widely used in the context of many-body systems, where the unfolding may be difficult to carry out properly.

\begin{figure}[b]
\includegraphics[scale=1.2]{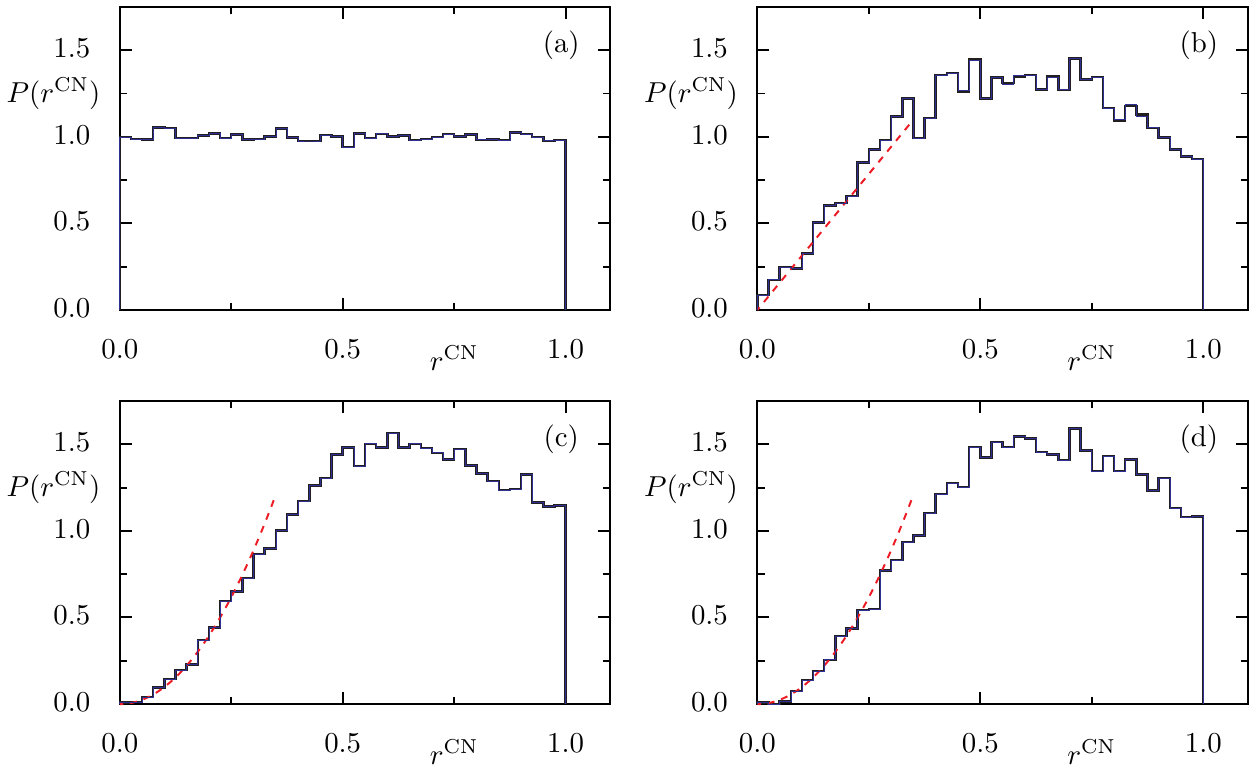}
\caption{Density of the ratios \CN/\SCN{} for
         (a) circle billiard,
         (b) cardioid billiard,
         (c) standard map, and
         (d) Riemann zeta function.
         The red dashed line in (b) shows the linear behaviour
         $\pi r^{\CN}$
         and in (c) and (d) the quadratic behaviour $\left (\pi r^{\CN}\right)^2$
         is shown.
         }
         \label{fig:ratio-CN-2CN}
\end{figure}

An alternative to the above measure with a cleaner interpretation in terms of the
ratio of the spacing of the two closest levels is given by
\begin{equation} \label{eq:r-i-CN}
  r_i^{\CN}  = \frac{s_i^{\CN}}{s_i^{\SCN}} .
\end{equation}
As for the ratio statistics, Eq.~\eqref{eq:ratio-statistics},
this quantity does not require an unfolding of the levels.
There are circumstances in a spectrum
for which the \FNtxt{} would be the \SCNtxt{} (especially for the GUE and GSE), and $r_i^\CN{}$
coincides with $r_i$,  but this does not always occur. The
distinctions between these closely related measures are nicely reflected in the
different densities, \FN{} and \SCN{}, of the previous section. Though
an analytical prediction for this quantity is not given, some
salient features can be quickly inferred from this
interpretation. For example, the occurrence of eigenvalues in
integrable systems is described by  a Poisson process which
means that the probability of
finding a level is independent of the presence or absence of another
level, therefore the probability of the ratio of $r_i^\CN$ would be equal
for all ratios. This is clearly seen in Fig.~\ref{fig:ratio-CN-2CN}~(a).
For the RMT ensembles, level repulsion implies zero probability
for the ratios $r^\CN$ at the origin.  To lowest order, the behaviour
near small ratios is $P(r^{\CN}) \propto  (r^{\text{\CN}})^\beta$.
These features are visible in the numerical histograms
of Fig.~\ref{fig:ratio-CN-2CN}~(b), (c) and (d).
The comparison of the mean ratios of $r$ and $r^\CN$ for the circle billiard, cardioid billiard,
standard map, and Riemann zeroes are given in Table~\ref{tab:mean-ratio}.
The increase of the mean value of $P(r^\CN)$ relative to $P(r)$
is consistent with the fact that the mean of $P_{\SCN}(s)$ is smaller than
$P_{\FN}(s)$ and both of them are continuous, smooth densities.
\begin{table}
\begin{center}
	\begin{tabular}{|l|l|l|}\hline
		& $\langle r \rangle$   
		& $\langle r^{\CN}\rangle$    \T\B     
		\\ \hline
		(a) Circle billiard       & 0.3855 & 0.4988 \\
		(b) Cardioid billiard     & 0.5305 & 0.5689 \\
		(c) Standard map          & 0.6013 & 0.6187 \\
		(d) Riemann zeta function & 0.6019 & 0.6197 \\\hline
	\end{tabular}
\end{center}
\caption{Numerical mean values $\langle r\rangle$ and
         $\langle r^{\CN} \rangle$ of
          \CN/\SCN{} and \CN/\FN{}, respectively.
}
\label{tab:mean-ratio}
\end{table}

\section{Summary and discussion}

In this paper, the \CNtxt{} and \FNtxt{}
spacing probability densities are introduced and predictions obtained
for the random matrix ensembles GOE, GUE, and GSE
based on a $3\times 3$ matrix modeling, and Poisson spectra.
The results are successfully compared with numerical
computations for the integrable circle billiard,
the fully chaotic cardioid billiard, the standard map
with chaotic dynamics and broken time reversal symmetry, and the zeros of the Riemann zeta function.
More generally, the class of \kCNtxt{} spacing
statistics are discussed, which are numerically found to be well described by
 Gaussians with mean increasing by half and variances depending on the case.
One application of the \CNtxt{} spacing density is in the context of
perturbation theory, and the \FNtxt{} sheds light on understanding
the ratio defined using successive level spacings.

In addition, the statistics of the ratio of the \CNtxt{}  spacing to that of
 the \SCNtxt{} is introduced as alternative measure for the
correlation properties of complex quantum spectra,
which does not require unfolding. This ratio has a
simpler interpretation.  Finding an analytical prediction for the probability density of the ratio
is an interesting task left for the future.
A straight-forward generalization, left for future work, would be to consider
the statistics of the ratios $s_i^{\kCN}/s_i^{(k+1)\CN}$,
which would give insight into longer range correlations.
It is noteworthy that a recent work \cite{TekBhoSan2018} explores
certain higher range spacing ratios of a kind that is not based on ordering.

\section*{Acknowledgements}

One of us (ST) gratefully acknowledges support for visits to MPIPKS, TU Dresden, and URegensburg.

\section*{References}

\bibliographystyle{iopart-num-mod.bst}
\bibliography{abbrevs,extracted,references}

\providecommand{\newblock}{}
\begin{thebibliography}{10}
\providecommand{\eprint}[2][]{\url{#2}}

\bibitem{GuhMueWei1998}
Guhr T, M\"uller-Groeling A and Weidenm\"uller H~A 1998 Random-matrix theories
  in quantum physics: common concepts {\em Phys.~Rep.\/} {\bf 299} 189--425

\bibitem{RMTScholarpedia}
Fyodorov Y 2011 {R}andom matrix theory {\em Scholarpedia\/} {\bf 6} 9886

\bibitem{BerTab1977}
Berry M~V and Tabor M 1977 Level clustering in the regular spectrum {\em
  Proc.~R.~Soc.~Lon.~A\/} {\bf 356} 375--394

\bibitem{BohGiaSch1984}
Bohigas O, Giannoni M~J and Schmit C 1984 Characterization of chaotic quantum
  spectra and universality of level fluctuation laws {\em Phys.~Rev.~Lett.\/}
  {\bf 52} 1--4

\bibitem{DysMeh1963}
Dyson F~J and Mehta M~L 1963 Random matrices and the statistical theory of
  energy levels {IV} {\em J.~Math.~Phys.\/} {\bf 4} 701--712

\bibitem{BroFloFreMelPanWon1981}
Brody T~A, Flores J, French J~B, Mello P~A, Pandey A and Wong S~S~M 1981
  Random-matrix physics: spectrum and strength fluctuations {\em
  Rev.~Mod.~Phys.\/} {\bf 53} 385--479

\bibitem{Ber1985}
Berry M~V 1985 Semiclassical theory of spectral rigidity {\em
  Proc.~R.~Soc.~Lon.~A\/} {\bf 400} 229--251

\bibitem{Ber1988}
Berry M~V 1988 Semiclassical formula for the number variance of the riemann
  zeros {\em Nonlinearity\/} {\bf 1} 399--407

\bibitem{LakSriKetBaeTom2016}
Lakshminarayan A, Srivastava S~C~L, Ketzmerick R, B{\"a}cker A and Tomsovic S
  2016 Entanglement and localization transitions in eigenstates of interacting
  chaotic systems {\em Phys.~Rev.~E\/} {\bf 94} 010205(R)

\bibitem{TomLakSriBae2018}
Tomsovic S, Lakshminarayan A, Srivastava S~C~L and B\"acker A 2018 Eigenstate
  entanglement between quantum chaotic subsystems: Universal transitions and
  power laws in the entanglement spectrum {\em Phys.~Rev.~E\/} {\bf 98} 032209

\bibitem{ForOdl1996}
Forrester P~J and Odlyzko A~M 1996 Gaussian unitary ensemble eigenvalues and
  {Riemann} $\zeta$ function zeros: {A} nonlinear equation for a new statistic
  {\em Phys.~Rev.~E\/} {\bf 54} R4493--R4495

\bibitem{HerOngUsaMatBar2007}
Herman D, Ong T~T, Usaj G, Mathur H and Baranger H~U 2007 Level spacings in
  random matrix theory and {Coulom}b blockade peaks in quantum dots {\em
  Phys.~Rev.~B\/} {\bf 76} 195448

\bibitem{BalHil1976}
Baltes H~P and Hilf E~R 1976 {\em Spectra of Finite Systems\/} (Mannheim, Wien,
  Z\"urich: Bibliographisches Institut)

\bibitem{OgaHus2007}
Oganesyan V and Huse D~A 2007 Localization of interacting fermions at high
  temperature {\em Phys.~Rev.~B\/} {\bf 75} 155111

\bibitem{AtaBogGirRou2013}
Atas Y~Y, Bogomolny E, Giraud O and Roux G 2013 Distribution of the ratio of
  consecutive level spacings in random matrix ensembles {\em
  Phys.~Rev.~Lett.\/} {\bf 110} 084101

\bibitem{AtaBogGirVivViv2013}
Atas Y~Y, Bogomolny E, Giraud O, Vivo P and Vivo E 2013 Joint probability
  densities of level spacing ratios in random matrices {\em J.~Phys.~A\/} {\bf
  46} 355204

\bibitem{ChaDeoKot2014}
Chavda N~D, Deota H~N and Kota V~K~B 2014 Poisson to {GOE} transition in the
  distribution of the ratio of consecutive level spacings {\em Phys.~Lett.~A\/}
  {\bf 378} 3012--3017

\bibitem{AleRig2014}
D'Alessio L and Rigol M 2014 Long-time behavior of isolated periodically driven
  interacting lattice systems {\em Phys.~Rev.~X\/} {\bf 4} 041048

\bibitem{KheChaKimSon2014}
Khemani V, Chandran A, Kim H and Sondhi S~L 2014 Eigenstate thermalization and
  representative states on subsystems {\em Phys.~Rev.~E\/} {\bf 90} 052133

\bibitem{Por1965}
{C E Porter (ed)} 1965 {\em Statistical Theories of Spectra: Fluctuations\/}
  (New York: Academic Press)

\bibitem{BohGiaSch1984b}
Bohigas O, Giannoni M~J and Schmit C 1984 Spectral properties of the
  {Laplacian} and random matrix theories {\em J.~Physique Lett.\/} {\bf 45}
  L1015--L1022

\bibitem{Wig1957}
Wigner E~P 1957 Results and theory of resonance absorption {\em Conference on
  Neutron Physics by Time-of-Flight\/} (Oak Ridge National Laboratory Report
  No. 2309) pp 59--70

\bibitem{Meh2004}
Mehta M~L 2004 {\em Random Matrices\/} 3rd ed (Elsevier Ltd.)

\bibitem{Mez2007}
Mezzadri F 2007 How to generate random matrices from the classical compact
  groups {\em Not.~Am.~Math.~Soc.\/} {\bf 54} 592--604

\bibitem{Sin1970}
Sinai {\relax Ya}~G 1970 Dynamical systems with elastic reflections {\em
  Russ.~Math.~Surv.\/} {\bf 25} 137--189

\bibitem{Bun1979}
Bunimovich L~A 1979 On the ergodic properties of nowhere dispersing billiards
  {\em Commun.~Math.~Phys.\/} {\bf 65} 295--312

\bibitem{Rob1983}
Robnik M 1983 Classical dynamics of a family of billiards with analytic
  boundaries {\em J.~Phys.~A\/} {\bf 16} 3971--3986

\bibitem{Woj1986}
Wojtkowski M 1986 Principles for the design of billiards with nonvanishing
  {L}yapunov exponents {\em Commun.~Math.~Phys.\/} {\bf 105} 391--414

\bibitem{Mar1988}
Markarian R 1988 Billiards with {Pesin} region of measure one {\em
  Commun.~Math.~Phys.\/} {\bf 118} 87--97

\bibitem{BaeDul1997}
B\"acker A and Dullin H~R 1997 Symbolic dynamics and periodic orbits for the
  cardioid billiard {\em J.~Phys.~A\/} {\bf 30} 1991--2020

\bibitem{Sza1992}
Sz\'asz D 1992 On the {$K$}-property of some planar hyperbolic billiards {\em
  Commun.~Math.~Phys.\/} {\bf 145} 595--604

\bibitem{Mar1993}
Markarian R 1993 New ergodic billiards: exact results {\em Nonlinearity\/} {\bf
  6} 819--841

\bibitem{LivWoj1995}
Liverani C and Wojtkowski M~P 1995 Ergodicity in {Hamiltonian} systems {\em
  Dynamics {{Reported}}\/} ({\em Dynamics Reported\/} no~4) ed Jones C~K~R~T,
  Kirchgraber U and Walther H~O ({Springer Berlin Heidelberg}) pp 130--202

\bibitem{CheHas1996}
Chernov N~I and Haskell C 1996 Nonuniformly hyperbolic {K}-systems are
  {Bernoulli} {\em Ergodic Theory Dynam.~Systems\/} {\bf 16} 19--44

\bibitem{PrivComProRob}
Prosen T and Robnik M private communication

\bibitem{Rob1984}
Robnik M 1984 Quantising a generic family of billiards with analytic boundaries
  {\em J.~Phys.~A\/} {\bf 17} 1049--1074

\bibitem{ProRob1993a}
Prosen T and Robnik M 1993 Energy level statistics in the transition region
  between integrability and chaos {\em J.~Phys.~A\/} {\bf 26} 2371--2387

\bibitem{BaeSteSti1995}
B\"acker A, Steiner F and Stifter P 1995 Spectral statistics in the quantized
  cardioid billiard {\em Phys.~Rev.~E\/} {\bf 52} 2463--2472

\bibitem{Chi1979}
{Chirikov} B~V 1979 {A universal instability of many-dimensional oscillator
  systems} {\em Phys.~Rep.\/} {\bf 52} 263--379

\bibitem{Gor2012}
Gorodetski A 2012 On stochastic sea of the standard map {\em
  Commun.~Math.~Phys.\/} {\bf 309} 155--192

\bibitem{BerBalTabVor1979}
Berry M~V, Balazs N~L, Tabor M and Voros A 1979 Quantum maps {\em
  Ann.~Phys.~(N.Y.)\/} {\bf 122} 26--63

\bibitem{HanBer1980}
Hannay J~H and Berry M~V 1980 Quantization of linear maps on a torus ---
  {F}resnel diffraction by a periodic grating {\em Physica~D\/} {\bf 1}
  267--290

\bibitem{ChaShi1986}
Chang S~J and Shi K~J 1986 Evolution and exact eigenstates of a resonant
  quantum system {\em Phys.~Rev.~A\/} {\bf 34} 7--22

\bibitem{KeaMezRob1999}
Keating J~P, Mezzadri F and Robbins J~M 1999 Quantum boundary conditions for
  torus maps {\em Nonlinearity\/} {\bf 12} 579--591

\bibitem{DegGra2003bwcrossref}
Degli~Esposti M and Graffi S 2003 Mathematical aspects of quantum maps {\em The
  Mathematical Aspects of Quantum Maps\/} ({\em Lect.~Notes Phys.\/} vol 618)
  (Berlin: Springer-Verlag) pp 49--90

\bibitem{Rie1859}
Riemann G~F~B 1859 Ueber die {Anzahl} der {Primzahlen} unter einer gegebenen
  {Gr\"osse} {\em Monatsber.~K\"onigl.~Preuss.~Akad.~Wiss.~Berlin\/}  671--680

\bibitem{Man1905}
v~Mangoldt H 1905 Zur {Verteilung} der {Nullstellen} der {Riemannschen}
  {Funktion} $\xi$(t) {\em Math.~Ann.\/} {\bf 60} 1--19

\bibitem{TitHea1988}
Titchmarsh E~C and Heath-Brown D~R 1988 {\em The Theory of the Riemann
  Zeta-Function\/} 2nd ed (Oxford University Press, New York)

\bibitem{Odl1987}
Odlyzko A~M 1987 On the distribution of spacings between zeros of the zeta
  function {\em Math.~Comp.\/} {\bf 48} 273--308

\bibitem{OdlTables}
Odlyzko A~M Tables of zeros of the {Riemann} zeta function
  {http://www.dtc.umn.edu/\~{}odlyzko/zeta\_tables/}

\bibitem{AurSte1990}
Aurich R and Steiner F 1990 Energy-level statistics of the
  {H}adamard-{G}utzwiller ensemble {\em Physica~D\/} {\bf 43} 155--180

\bibitem{French78}
French J~B, Mello P~A and Pandey A 1978 Statistical properties of many-particle
  spectra. {II}. two-point correlations and fluctuations {\em Ann.~Phys.\/}
  {\bf 113} 277--293

\bibitem{AurBaeSte1997}
Aurich R, B\"acker A and Steiner F 1997 Mode fluctuations as fingerprints of
  chaotic and non-chaotic systems {\em Int.~J.~Mod.~Phys.~B\/} {\bf 11}
  805--849

\bibitem{BapBeg1989}
Bapat R~B and Beg M~I 1989 Order statistics for nonidentically distributed
  variables and permanents {\em Sankhy\=a: The Indian Journal of Statistics,
  Series A\/} {\bf 51} 79--93

\bibitem{TekBhoSan2018}
Tekur S~H, Bhosale U~T and Santhanam M~S 2018 Higher-order spacing ratios in
  random matrix theory and complex quantum systems {\em Phys.~Rev.~B\/} {\bf
  98} 104305

\end{thebibliography}

\end{document}